\documentclass[12pt,a4paper]{revtex4}

\usepackage{amsmath}
\usepackage{graphicx}
\begin{document}

\begin{abstract}
We address a problem of generating a robust entangling gate between electronic and nuclear spins in the system of a single nitrogen-vacany centre coupled to a nearest $^{13}$Carbon atom in diamond against certain types of systematic errors  such as pulse-length and off-resonance errors. We analyse the robustness of various control schemes: sequential pulses, composite pulses and numerically-optimised pulses. We find that numerically-optimised pulses, produced by the gradient ascent pulse engineering algorithm (GRAPE), are more robust than the composite pulses and the sequential pulses. The optimised pulses can also be implemented in a faster time than the composite pulses.
\end{abstract}

\title{Robust control of entanglement in a Nitrogen-vacancy centre coupled to a $^{13}$C-Carbon nuclear spin in diamond}

\author{R. S. Said and J. Twamley}
\affiliation{\footnotesize Research Centre for Quantum Science \& Technology (QSciTech), and Department of Physics \& Electronic Engineering, Faculty of Science, Macquarie University, Sydney NSW 2109 Australia.}

\maketitle

\section{Introduction}

Solid-state technologies hold great promise towards the fabrication of large scale quantum devices. Hence, theoretical and experimental investigations into the control of quantum information in such systems have progressed quite rapidly and candidate technologies include   Phosphorus donor electrons in Silicon nanostructures \cite{solidQC, DonorQC}, Gallium Arsenide quantum dots \cite{QuantumDotsQC}, superconducting single-Cooper-pair boxes \cite{Nakamura},  circuit quantum electrodynamics \cite{cQED}, and single nitrogen-vacancy (NV) centres in diamond \cite{NVScience}.
A system based on NV centres in diamond is attractive because at room temperature, the NV centre displays strong spin polarisation under optical pumping, exhibits remarkable photostability and shows high fluorescence quantum yield \cite{NVQCreview}. The electronic spins of the NV centre are initialised and measured by optical means and can be manipulated by a microwave radiation. Through a hyperfine coupling, an interaction between the electronic spins of the NV centre and a nuclear spin of a nearby $^{13}$C atom can be  exploited to encode two-qubits and quantum logic can be executed via the application on microwave and radio-frequency radiation.
Oservations of Rabi oscillations of the single electronic spin \cite{exper1}, and the single nuclear spin \cite{exper2} in the NV centre, performed by optically detected magnetic resonance (ODMR) techniques, have paved the way for a realisation of the NV centre based quantum computer \cite{AP}. A more advanced quantum information processing (QIP) task, that is a demonstration of three-qubit entanglement has been recently reported in the literature \cite{experGHZ}. This latest development points to possibilities of multi-qubit QIP implementations. However, the use of accidental nearby nuclear spins poses some difficulties towards scaling up this technology. 

Because the qubit system interacts with an environment, QIP is subject to unavoidable errors. Hence, error control and avoidance schemes are necessary to achieve  reliable quantum computation. The errors can occur either in a random or systematic fashion. The random errors are due to decoherence processes while systematic errors occur when the physical apparatuses controlling the dynamics of the system operate in an imprecise but reproducible manner. In the spin system driven by microwave or radio-frequency radiation, non ideal values of the radiation properties (i.e. amplitude, phase, duration and frequency), cause such systematic errors. 
A common type of systematic error, namely {\it pulse-length error} (PLE), occurs when the radiation (assumed to be a rectangular pulse), is resonant with the target spin transition but whose application-time or amplitude differs in an unknown (but fixed in time) quantity from the ideal value. Another type of systematic error is {\it off-resonance error} (ORE). Off-resonance errors arises when the frequency of the control radiation is unexpectedly not on resonance with the spin transition and spin dynamics proceeds with an unknown (but fixed), detuning parameter.

In this paper, we describe schemes of generating an entangling gate between the electronic and nuclear spins in the NV centre in diamond which are robust against PLE and ORE systematic errors. In this letter, we investigate effects of such errors in a sequential application of rectangular pulses of microwave and radio-frequency. We compare the fidelity of such pulses with their more robust counterparts: composite pulses, and numerically derived GRAPE pulses \cite{grape}.  The sequential pulses have previously been used to generate two-qubit entanglement \cite{NVQCreview}, while composite pulses are known to be capable of correcting systematic errors in nuclear magnetic resonance (NMR) experiments \cite{composite}. 
In the next section, we describe the Nitrogen-Vacancy (NV) system and introduce the sequential pulses required to create a particular entangled state in this system.  Section III examines the decrease in gate fidelity due to systematic errors when we adopt the sequential, robust composite and robust GRAPE pulse control.
Results of the numerical simulations and performance comparisons between the composite and GRAPE pulses against PLE and ORE systematic errors are discussed in Section IV, together with conclusions. We find that GRAPE pulses are more robust than the sequential and composite pulses, and faster than the composite pulses.

\section{Entanglement generation}
To describe the NV, we follow a model described in detail in \cite{AP} and concentrate our discussions on the fine and hyperfine structure of the ground state of the NV centre coupled to the $^{13}$C atom. 
\begin{figure}[bt]
\begin{center}
\setlength{\unitlength}{.55cm}
\begin{picture}(7,10)
\put(-1,0.5){\includegraphics[width=8.4\unitlength]{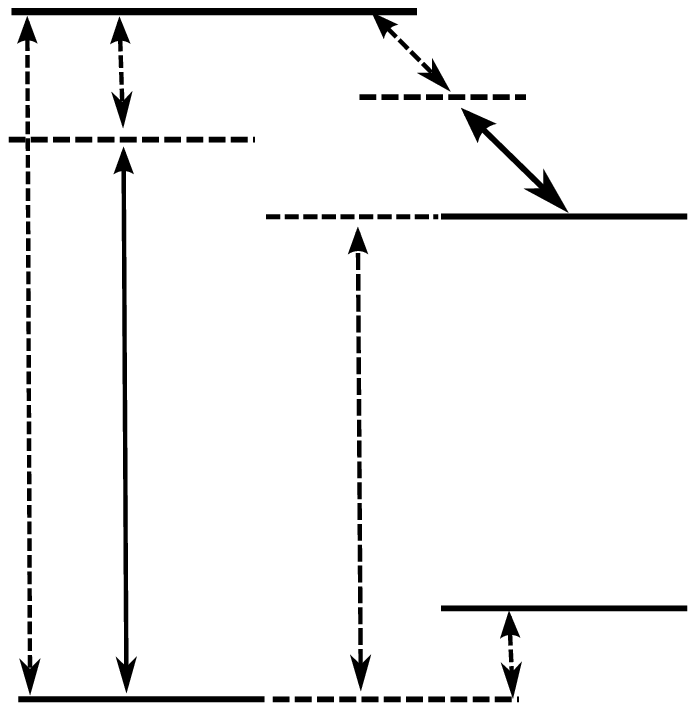}}
\put(-5.9,.4){$|0\rangle \equiv |0\rangle_e \otimes |0\rangle_n $}
\put(-5.9,8.5){$|2\rangle\equiv |1\rangle_e \otimes |0\rangle_n$}
\put(-1.9,4.8){$\omega_{02}$}
\put(-6,4.8){$\mathbf{^3 A}$}
\put(.7,4.2){$\omega_{m}$}
\put(.7,8.0){$\delta_{m}$}
\put(2.3,3.8){$\omega_{03}$}
\put(7.5,1.5){$|1\rangle\equiv |0\rangle_e \otimes |1\rangle_n$}
\put(7.5,6.4){$|3\rangle\equiv |1\rangle_e \otimes |1\rangle_n$}
\put(4.1,1.05){$\omega_{01}$}
\put(5.8,7.2){$\omega_{r}$}
\put(4.4,8.4){$\delta_{r}$}
\end{picture}
\end{center}
\caption{{Energy level diagram of the ground state $^3$A of the NV centre in the absence of an external magnetic field. The $|0\rangle - |1\rangle$, and $|2\rangle - |3\rangle$, enery splitings are due to the hyperfine interaction between the NV electron and the nuclear spin of the $^{13}$Carbon atom. The $|0\rangle - |2\rangle$, $|2\rangle - |3\rangle$ transitions are driven by MW, RF radiation observing selection rules.
}
\label{energylevel}}
\end{figure}
When there is no external magnetic field aligned with the quantisation axis of the NV centre, the system has four levels in the $^3$A manifold as depicted by Figure 1. This is due to the degeneracy of the $m_s=\pm1$ electronic spins $\left(S=1\right)$, of the centres and the interaction with the nuclear spin $\left(I=1/2\right)$, of the Carbon atom. 
For clarity and consistency, we adopt the same notations for the four spin levels as those used in \cite{AP}:
\begin{eqnarray}
|0\rangle &=& |0\rangle_e \otimes |0\rangle_n = \frac{1}{\sqrt{2}} \left( |\uparrow\downarrow\rangle +  |\downarrow\uparrow\rangle \right)_e \otimes |\uparrow\rangle_n, \\
|1\rangle &=& |0\rangle_e \otimes |1\rangle_n =\frac{1}{\sqrt{2}} \left( |\uparrow\downarrow\rangle +  |\downarrow\uparrow\rangle \right)_e \otimes |\downarrow\rangle_n, \\
|2\rangle &=& |1\rangle_e \otimes |0\rangle_n =  \frac{1}{\sqrt{2}} \left( |\uparrow\uparrow\rangle +  |\downarrow\downarrow\rangle \right)_e   \otimes |\uparrow\rangle_n, \\
|3\rangle &=& |1\rangle_e \otimes |1\rangle_n = \frac{1}{\sqrt{2}} \left( |\uparrow\uparrow\rangle +  |\downarrow\downarrow\rangle \right)_e \otimes |\downarrow\rangle_n.
\end{eqnarray}
The states $\left\{|0\rangle_e,|1\rangle_e\right\}$ are eigenstates associated with the electronic spins $\left\{m_s=0, m_s=- 1\right\}$ while the states $\left\{ |\uparrow\rangle_n,|\downarrow\rangle_n \right\}$ correspond to the nuclear spins $\left\{m_I=-1/2, m_I=+ 1/2\right\}$. From \cite{AP}, the $|0\rangle-|2\rangle$ and $|2\rangle-|3\rangle$ transitions are found to be $\omega_{02}\approx 2.88$ GHz and $\omega_{03}\approx 130$ MHz respectively, while $\omega_{01}\approx 2$ MHz. 
In general, the two frequencies $\omega_m$ and $\omega_r$ can be set slightly off resonance as parameterised by detunings $\delta_m$ and $\delta_r$. These detunings have unknown but constant values in the case of ORE.

The Hamiltonian of the system described in Figure 1 can be expressed as
\begin{eqnarray}
\hat H &=& \omega_{02} \hat\sigma_{22}+\omega_{03} \hat\sigma_{33} + \omega_{01} \hat\sigma_{11} \nonumber \\
&&
-\frac{1}{2}\left( \Omega_m e^{i \omega_m t}\hat\sigma_{20} 
+ \Omega_r e^{i\omega_r t}\hat\sigma_{23} + H.c. \right),
\end{eqnarray}
where $\hat\sigma_{pq}=|p\rangle\langle q |$, $\Omega_{m}(\Omega_{r})$ are the Rabi frequencies driving the MW (RF) transitions. The first line of (5) is the self-energy of the system relative to the ground state $|0\rangle$, while the reminder of the Hamiltonians describes the interaction of the external radiation with the system. We now move to the interaction picture (IP) defined by $\hat U_0(t)\equiv\exp(-i \hat H_0t)$, with
$\hat H_0 = a \hat\sigma_{22} + b \hat\sigma_{33} + c \hat\sigma_{11} + d \hat\sigma_{00},
$,
to obtain $\hat H_{eff}=\hat U \hat H \hat U^\dag - i \hat U^\dag\dot{\hat U}$,
\begin{eqnarray}
\hat H_{eff} &=& 
\left(\omega_{02} - a\right) \hat\sigma_{22}
+ \left(\omega_{03} - b\right) \hat\sigma_{33} \nonumber\\
&&+ \left(\omega_{01} - c\right) \hat\sigma_{11}
- d \hat\sigma_{00} \nonumber\\
&& - \frac{1}{2} \left( \Omega_m  \hat\sigma_{20} +
 \Omega_r  \hat\sigma_{23} + H.c. \right),
\end{eqnarray}
where we have set $a-b=\omega_r,\, a-d=\omega_m$.
Defining
$
\hat I = \hat\sigma_{00} + \hat\sigma_{22} + \hat\sigma_{33} + \hat\sigma_{11}, \,
\hat\sigma_z^{20} = - \hat\sigma_{00} + \hat\sigma_{22}, \,
\hat\sigma_z^{23} = \hat\sigma_{22} - \hat\sigma_{33}, \,
\hat I^{31} = \hat\sigma_{33} + \hat\sigma_{11},
$, 
gives the effective Hamiltonian a new form expressed as
\begin{eqnarray}
\hat H_{eff} &=& 
  \frac{1}{2} \left(3\omega_m -2\delta_m -\delta_r -\omega_{01} - 3a +c \right) \hat I \nonumber\\
  &&
+ \frac{1}{2} \left(\omega_m +2\delta_m -\delta_r -\omega_{01} -a +c \right) \hat \sigma_z^{20} \nonumber\\
&&
+ \left(-\omega_m -\delta_m +\delta_r +\omega_{01} +a -c \right) \hat \sigma_z^{23} \nonumber\\
&&
+ \frac{1}{2} \left(-3\omega_m -2\delta_m +\delta_r +3\omega_{01} + 3a - 3c  \right) \hat I^{31} \nonumber\\
&& - \frac{1}{2} \left( \Omega_m  \hat\sigma_{20} + H.c. \right)
- \frac{1}{2} \left( \Omega_r  \hat\sigma_{23} + H.c. \right).
\end{eqnarray}
The terms having $\hat I$ and $\hat I^{31}$ in the above expression vanish by taking
$
a=\omega_m + \frac{2}{3} \delta_m - \frac{1}{3} \delta_r,\; c=\omega_{01}, 
$, 
so  that,
\begin{eqnarray}
\hat{H}_{eff} &=& \frac{1}{3} \delta \left( \hat\sigma_z^{20} + \hat\sigma_z^{23} \right) \nonumber\\
&& - \frac{1}{2} u_m \left( \cos \theta_m  \hat\sigma_x^{20} + \sin \theta_m  \hat\sigma_y^{20} \right) \nonumber \\
&&- \frac{1}{2} u_r \left( \cos \theta_r  \hat\sigma_x^{23} + \sin \theta_r  \hat\sigma_y^{23} \right),\label{eq12}
\end{eqnarray}
where express the control pulse $\Omega_{m,r}=u_{m,r}\exp\left(i \theta_{m,r}\right)$ and we have chosen $\delta_m=\delta_r=\delta$. 
We note that $u_{m,r}$ and $\theta_{m,r}$ are a real, and describe the control amplitudes and control phases, and  use $\hat\sigma_x^{pq}=\hat\sigma_{pq}+\hat\sigma_{qp}$, $\hat\sigma_y^{pq}=i\left(\hat\sigma_{pq}-\hat\sigma_{qp}\right)$.
We can clearly see from (\ref{eq12}) that the Hamiltonian of the system is reduced effectively from the four-state system to a three-state system involving only the states $|0\rangle$, $|2\rangle$ and $|3\rangle$, which can be expressed as
\begin{equation}
\hat{H}_{eff} =
- \frac{1}{2}
\left(
\begin{array}{ccc}
\frac{2}{3}\delta & u_m e^{-i\theta_m} & 0 \cr
u_m e^{i\theta_m} & -\frac{4}{3}\ \delta & u_r e^{i\theta_r} \cr
0 & u_r e^{-i\theta_r} & \frac{2}{3}\delta
\end{array}
\right).
\end{equation}
We now consider the sequential application of MW and RF unitaries via rectangular control pulses. We wish to generate the entangling gate $|\Psi_f\rangle = \hat R_{23} (\pi) \hat R_{02} (\pi/2) |0\rangle = \hat U_r \hat U_m |0\rangle$. We consider the resonant case ($\delta=0$), and we first apply a MW pulse on the $|0\rangle \leftrightarrow |2\rangle$ transition for time $t_m=\frac{\pi}{2} u_m^{-1}$ to get 
$\hat{U}_m = e ^{\frac{1}{4} i \hat\sigma_y^{20}},
$,  
and 
$| \psi_H \rangle = \hat{U}_m |0\rangle = \frac{1}{\sqrt{2}} (|0\rangle - |2\rangle)
$. 
Switching off the microwave radiation and subsequently applying a radio-frequency pulse with a phase of $\theta_r=\frac{\pi}{2}$, for a duration of $t_r=\pi u_r^{-1}$ on resonance with the $|2\rangle\leftrightarrow|3\rangle$ transition produces the gate 
$\hat{U}_r= e^{  \frac{\pi}{2} i \hat\sigma_y^{23}},
$ 
which transforms the state $|\psi_m\rangle$ into another superposition state $|\psi_{B}\rangle$,
\begin{eqnarray}
|\psi_B \rangle= \hat{U}_r |\psi_H \rangle= \hat{U}_r \hat{U}_m |0\rangle = \frac{1}{\sqrt{2}} \left(|0\rangle -|3\rangle\right)=  \frac{1}{\sqrt{2}} \left(|0,0\rangle -|1,1\rangle\right).
\end{eqnarray}
The state $|\psi_{B}\rangle$ is an entangled Bell state in the coupled system of electronic and nuclear spins. The sequential application of the on-resonance microwave and radio-frequency pulses generates a sequential unitary gate $\hat{U}_{sq}$, 
\begin{eqnarray}
\hat{U}_{sq} = \hat{U}_r \hat{U}_m = \frac{1}{\sqrt{2}}\left(
\begin{array}{ccc}
1 & 1 & 0 \cr
0 & 0 & -\sqrt{2} \cr
-1 & 1 & 0
\end{array}
\right),
\end{eqnarray}
which takes a total time $t_{sq}=t_m+t_r=\frac{3\pi}{2} u_m^{-1}$, if the amplitudes $u_m=u_r$. 

\section{Systematic errors}
To study the effects of systematic errors on the gate fidelity, we first quantify the PLE and ORE by two error fractions $\epsilon_f=(T'-T)/T$, and $\epsilon_g=\delta/\Lambda$, respectively, where $-1\leq\{\epsilon_f,\epsilon_g\} \leq 1$. We use $T$ and $T'$ to denote the ideal and non-ideal pulse application times and $\Lambda$ is taken to be the fixed maximum amplitude of the microwave or radio-frequency pulses. 
In the presence of off-resonance error we consider the case when in the sequential pulse, $\max\{u_m\}=\max\{u_r\}=\Lambda$. 

When the sequential pulse suffering from pulse-length error is applied to the system the actual gate executed is highly dependent on the error fraction $\epsilon_f$, and we can write,
\begin{eqnarray}
\hat{U}_{sq}^f= \hat{U}_{r}^f \hat{U}_{m}^f = e^{  \pi/2 (1+\epsilon_f)  i \hat{\sigma}_y^{23}} e^{  \pi/4 (1+\epsilon_f) i \hat{\sigma}_y^{20}}. \label{Usqf}
\end{eqnarray}
We use the standard gate overlap fidelity $F$ to measure the closeness between the generated gate, say $\hat{U}_{a}$, and the target gate $\hat{U}_{i}$ \cite{grape2},
\begin{eqnarray}
F= \left| {\frac{\mathbf{Tr} \left(\hat{U}_{a}^\dag \hat{U}_{i}\right) }{\mathbf{Tr} \left(\hat{U}_{i}^\dag \hat{U}_{i}\right) }}\right| ^{\frac{1}{2}}. \label{FidelityEq}
\end{eqnarray}
One can show that the fidelity of the actual gate $\hat{U}_{sq}^f$ with respect to the ideal one $\hat{U}_{sq}$ is
\begin{eqnarray}
F_{sq}^f =
\left| {\frac{\mathbf{Tr} \left(\hat{U}_{sq}^{f\,\dag} \hat{U}_{sq} \right)}{\mathbf{Tr} \left(\hat{U}_{sq}^\dag \hat{U}_{sq} \right)}}\right| ^{\frac{1}{2}} \approx 1-{\frac {5{\pi }^{2}}{96}}{\epsilon_f}^{2}+{\frac {{\pi }^{4}}{4608}}{\epsilon_f}^{4}, \label{21}
\end{eqnarray}
and is numerically plotted in Figure 3. 
The quadratic term of $\epsilon_f$ in (\ref{21}) significantly reduces the fidelity. 
Hence, It is very desirable to find other types of control pulse that can suppress and possibly eliminate the quadratic term of $\epsilon_f$ from the gate fidelity.

In the presence of the off-resonance error, the actual gate implemented by the sequential pulse is
\begin{eqnarray}
\hat{U}_{sq}^g(\epsilon_g)=\hat{U}_{r}^g(\epsilon_g) \hat{U}_{m}^g(\epsilon_g),
\end{eqnarray}
where 
\begin{eqnarray}
\hat{U}_{r}^g &=& e^{  - \frac{\pi}{3} i \epsilon_g (\hat{\sigma}_z^{20} + \hat{\sigma}_z^{23})
+ \frac{\pi}{2}   i \hat{\sigma}_y^{23}}, \\
\hat{U}_{m}^g &=& e^{  - \frac{\pi}{6} i \epsilon_g (\hat{\sigma}_z^{20} + \hat{\sigma}_z^{23}) + \frac{\pi}{4}  i \hat{\sigma}_y^{20}}.
\end{eqnarray}
Due to the complexity of the analytical expression, the fidelity of the gate $\hat{U}_{sq}^g$ is calculated numerically and plotted in Figure 5.

\subsection{Composite pulses}
Originally invented in NMR by Levitt and Freeman \cite{Levitt}, composite pulses offset the effect of systematic errors by replacing single quantum operations with several quantum operations which are designed to cancel out systematic errors. 
Composite pulses have been developed to implement robust arbitrary single qubit gate operations \cite{composite,compositeNJP}. 
In our case, we apply two particular types of composite pulses, (A) Broad Band Number 1 (BB1) composite pulses to correct for PLEs and (B) compensation for off-resonant errors with two CORPSE pulses.
To apply these composite pulses, we first define
generalised forms of the pulses subject to imperfect microwave and radio-frequency controls. In the presence of pulse-length error, these generalised forms are
\begin{eqnarray}
&\hat{{\cal U}}_m^f (\tau_{m},\theta_{m}) =
e^{i \frac{1}{2} (\cos\theta_m \hat{\sigma}_x^{20} + \sin\theta_m \hat{\sigma}_y^{20}) u_m (1-\epsilon_f) \tau_m}, &\\
&\hat{{\cal U}}_r^f \left(\tau_{r},\theta_{r} \right) =
e^{i \frac{1}{2} (\cos\theta_r \hat{\sigma}_x^{23} + \sin\theta_r \hat{\sigma}_y^{23}) u_r (1-\epsilon_f) \tau_r}.&
\end{eqnarray}
On the other hand, the generalised quantum operations in the case of off-resonance errors are
\begin{eqnarray}
\hat{{\cal U}}_m^g (\tau_{m},\theta_{m}) =
e^{- i ( \frac{i \epsilon_g}{3}\hat{Z} - \frac{1}{2} ( \cos\theta_m \hat{\sigma}_x^{20} + \sin\theta_m \hat{\sigma}_y^{20}) ) u_m \tau_m},  \\
\hat{{\cal U}}_r^g (\tau_{r},\theta_{r}) =
e^{- i ( \frac{i \epsilon_g}{3}\hat{Z} - \frac{1}{2} ( \cos\theta_m \hat{\sigma}_x^{23} + \sin\theta_m \hat{\sigma}_y^{23}) ) u_r \tau_r},
\end{eqnarray}
where $\hat{Z}=\hat{\sigma}_z^{20} + \hat{\sigma}_z^{23}$. 

We formulate the composite pulse analogue of the sequential gate (\ref{Usqf}) by replacing $\hat{U}_m^f$ and $\hat{U}_r^f$ with their composite counterparts.
For pulse-length errors we use BB1 composite pulses to replace $U_m^f$ and $U_r^f$ by $\hat{{\cal C}}^f_m$ and $\hat{{\cal C}}^f_r$ \cite{composite},
\begin{widetext}
\begin{eqnarray}
\hat{\cal C}^f_m &=& 
\hat{\cal U}_m^f \left(\pi/4,\pi/2\right)
\hat{\cal U}_m^f \left(\pi,1.04 \pi \right)
\hat{\cal U}_m^f \left(2\pi,2.12\pi \right)
\hat{\cal U}_m^f \left(\pi,1.04 \pi \right)
\hat{\cal U}_m^f \left(\pi/4,\pi/2\right), \\
\hat{\cal C}^f_r &=& 
\hat{\cal U}_r^f \left(\pi/2,\pi/2 \right)
\hat{\cal U}_r^f \left(\pi, 1.08\pi\right)
\hat{\cal U}_r^f \left(2\pi,  2.24\pi\right)
\hat{\cal U}_r^f \left(\pi, 1.08\pi\right)
\hat{\cal U}_r^f \left(\pi/2,\pi/2 \right),
\end{eqnarray}
\end{widetext}
where $\tau_m$ and $\tau_r$ are expressed in the units of $u_m^{-1}$ and $u_r^{-1}$. 
One can verify that in the absence of PLE, we have 
$
\hat{\cal C}^f_r (\epsilon_f=0) \hat{\cal C}^f_m (\epsilon_f=0) = \hat{U}_{sq}.
$ 
The composite pulses $\hat{\cal C}^f=\hat{\cal C}^f_r \hat{\cal C}^f_m$, takes a total time $\tau^f = 9\pi u_m^{-1}$ when both real amplitudes $u_m=u_r$ are the same. The total time is exactly 6 times longer than that of the sequential pulse. The fidelity of the BB1 composite gate, $\hat{\cal C}^f$, plotted in the Figure 3, is numerically calculated through the following equation,
\begin{eqnarray}
F_{c}^f = \left| {\frac{\mathbf{Tr} \left(\hat{\cal C}^{f\,\dag} \hat{U}_{sq}\right) }{\mathbf{Tr} \left(\hat{U}_{sq}^\dag \hat{U}_{sq} \right)}}\right| ^{\frac{1}{2}}.
\end{eqnarray}
The range of good fidelity (defined for $F\geq 0.9$) has expanded. In contrast with the sequential gate, where the good fidelity holds approximately only for  $|\epsilon_f| < 0.4$, the BB1 composite gate maintains the good fidelity for $|\epsilon_f| < 0.7$.

In the case of ORE, we replace the gates $\hat{U}_m^g$ and $\hat{U}_r^g$, with these CORPSE counterparts \cite{compositeNJP},
\begin{eqnarray}
\hat{\cal C}^g_m =
\hat{\cal U}_m^g (2.14\pi,\frac{\pi}{2} )
\hat{\cal U}_m^g (1.77\pi,-\frac{\pi}{2}  )
\hat{\cal U}_m^g (0.14\pi,\frac{\pi}{2}  ), \\
\hat{\cal C}^g_r =
\hat{\cal U}_r^g (7\pi/3,\frac{\pi}{2} )
\hat{\cal U}_r^g (5\pi/3, - \frac{\pi}{2} )
\hat{\cal U}_r^g (\pi/3, \frac{\pi}{2} ),
\end{eqnarray}
to produce the complete gate $\hat{\cal C}^g=\hat{\cal C}^g_r \hat{\cal C}^g_m $, which takes a total time of $\tau^g\approx 5.59 \times t_{sq}$, nearly six times longer than the sequential pulses assuming $u_m=u_r$. We plot the fidelity of the gate numerically calculated based on Equation (\ref{FidelityEq}) in Figure 5, and find that the CORPSE pulses do not correct the off-resonance error. This is due to the Autler-Townes splitting of the NMR transition due to the ESR excitation of the MW transition \cite{Wei1999}. Thus one cannot use the naive CORPSE pulse in each of the ESR and NMR sequential pulses to correct for ORE.

\subsection{Robust GRAPE Pulses}
We now explore whether one can obtain robust operations in a shorter time than the composite pulses by simultaneous irradiations both the MW and RF transitions. Rapid and robust control quantum control is important as the total gate duration significantly determines the number of quantum operations that can be performed before decoherence degrades the quantum coherences. Initially proposed for NMR experiments, the GRAPE algorithm produce pulses that minimise the time required to implement a target unitary operator \cite{grape} and some quantum algorithmic elements \cite{grape2}.
GRAPE pulses have been experimentally demonstrated in a single qubit trapped ion system \cite{ion}. A GRAPE based scheme is also proposed to control a coupled Josephson qubit system \cite{Josephson}, and later extended to control open quantum systems in the Markovian domain \cite{Markov}. It has also been applied in implementing high-fidelity single qubit operations in a noisy environment due to random telegraph noise in superconducting solid-state qubits \cite{Mikko}. 

We briefly summarise how the GRAPE algorithm works \cite{grape} and how it can achieve time-optimal control and robustness against systematic errors.
We start by writing the unitary evolution under the Hamiltonian (\ref{eq12}) in the form of 
\begin{eqnarray}
\hat U = {\cal T} \int e^{-i \left( \hat H_s + \sum_{k=1}^4 u_k(t) \hat{H}_k \right) t} dt, \label{Udes}
\end{eqnarray}
where $u_k(t)$ and $\hat H_k$ are the control pulses and control Hamiltonians, expressed as follows
\begin{eqnarray}
u_1(t) &=- \frac{u_m(t)}{2} \cos \theta_m(t),\qquad
u_2(t) &=- \frac{u_m(t)}{2} \sin \theta_m(t),\cr
u_3(t) &= - \frac{u_r(t)}{2} \cos\theta_r(t),\qquad
u_4(t) &= - \frac{u_r(t)}{2} \sin\theta_r(t),
\end{eqnarray}
and
$\{\hat{H}_1,\hat{H}_2,\hat{H}_3,\hat{H}_4\}=\{\hat\sigma_x^{20}, \hat\sigma_x^{20}, \hat\sigma_x^{23}, \hat\sigma_y^{23}\}$,
and $\hat H_s$ is a drift Hamiltonian, 
$
\hat H_s =  \frac{1}{3} \delta \left( \hat\sigma_z^{20} + \hat\sigma_z^{23} \right) = \frac{1}{3} \delta \hat Z,
$
and with the initial condition, $\hat{U}(t=0)=\hat{I}$. 
We wish to numerically optimise the controls $u_k(t)$ in a particular application time $t=T$, such that $\hat{U}(T)$ approaches a target gate $\hat{U}_T$. This is equivalent to maximising a performance function $P$,
\begin{eqnarray}
{P} = \left| \mathbf{Tr}\,  \hat{U}_T^\dag \hat{U}(T) \right|^2.
\end{eqnarray} 
One considers discretising time $\Delta t = T/N$, where $N$ is a number of time steps/bins, and during each time bin the control $u_k$'s are constant. We can approximate (\ref{Udes}),
\begin{eqnarray}
\hat{U} = \hat{U}_N \hat{U}_{N-1} ... \hat{U}_1, \label{Uapprox}
\end{eqnarray}
which gives an approximation of the performance function to be
\begin{eqnarray}
{P} &=& \mathbf{Tr}\, (\hat{U}_T^\dag \hat{U}_N \hat{U}_{N-1} ... \hat{U}_1) \times    \cr &&    \mathbf{Tr} \, (( \hat{U}_1 ... \hat{U}_{N-1} \hat{U}_N)^\dag \hat{U}_T).
\end{eqnarray}
From \cite{grape}, the gradient $g(j)=\delta{P}/\delta u_k(j)$, to ${\cal O}(\Delta t)$, is written as
\begin{eqnarray}
g(j) =  -2\mathbf{Re}(\mathbf{Tr} ( i \Delta t \hat A_j^\dag\hat{H}_k \hat B_j) \mathbf{Tr} (\hat B_j^\dag \hat A_j) ),
\end{eqnarray}
where $\hat A_j=\hat{U}_{j+1}^\dag ... \hat{U}_{N}^\dag \hat{U}_T$ and $\hat B_j=\hat{U}_j... \hat{U}_1$, and the performance function ${P}$ always increases if we update
\begin{eqnarray}
u_k(j) \rightarrow u_k(j) + \epsilon g(j),
\end{eqnarray}
where a small step size $\epsilon$ is used. It is also necessary to add an additional term $g_{max} (j)$,
\begin{eqnarray}
g_{max} (j)=-2 \alpha_p u_k(j) \Delta t,
\end{eqnarray}
to the gradient $g(j)$, to penalise excessive microwave or radio-frequency power.

The pulse-length and off-resonance errors are incorporated into the algorithm by replacing the ideal $\hat U_j$ in Equation (\ref{Uapprox}) with $\hat U_j^f$ and $\hat U_j^g$ written as
\begin{eqnarray}
\hat{U}_j^{\epsilon_f} = e^{- i \Delta t (1-\epsilon_f)(\sum_{k=1}^4 u_k(j) \hat{H}_k )},\\
\hat{U}_j^{\epsilon_g} = e^{- i \Delta t ( \epsilon_g \hat{H}_s + \sum_{k=1}^4 u_k(j) \hat{H}_k )}.
\end{eqnarray}
The performance function to be optimised is the average performance function over set of error fractions, and defined as
\begin{eqnarray}
{P}^{f,g} &=& \frac{1}{N(\vec{\epsilon}_{f,g})} \sum_{\min\{ \vec{\epsilon}_{f,g}\}}^{\max \{ \vec{\epsilon}_{f,g} \}} \mathbf{Tr} (\hat{U}_T^\dag \hat{U}_N^{\epsilon_{f,g}}\hat{U}_{N-1}^{\epsilon_{f,g}} ... \hat{U}_1^{\epsilon_{f,g}}) \times \cr && \mathbf{Tr} (( \hat{U}_1^{\epsilon_{f,g}} ... \hat{U}_{N-1}^{\epsilon_{f,g}} \hat{U}_N^{\epsilon_{f,g}} )^\dag \hat{U}_T),
\end{eqnarray}
for the case of PLE or ORE (denoted by superscripts $f$ or $g$), where $N(\vec{\epsilon}_{f,g})$ is a number of elements in the set of error fractions $\vec{\epsilon}_{f,g}$ we wish to optimize over. This modification allows us to optimise the control pulses $u_k$ only for a certain range of error fractions, for example $-0.2\leq \epsilon_{f,g} \leq 0.2$. However, the real performance of the optimised pulse is checked through the gate fidelity as defined by Equation (\ref{FidelityEq}).

\section{Discussions and Conclusions}

\begin{figure}[ht]
\includegraphics[scale=0.65]{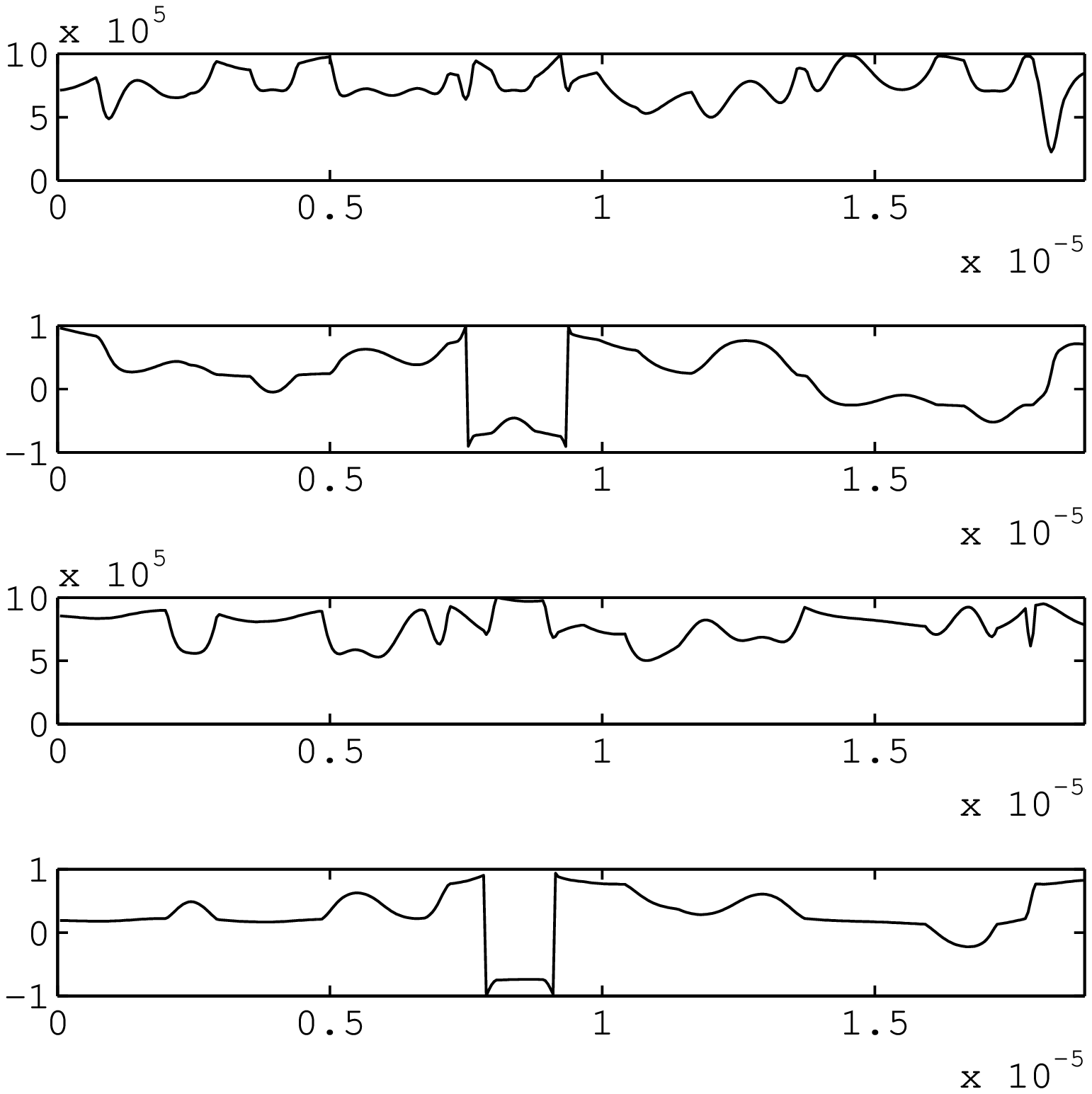}
\put(-180,10){Time (s)}
\put(-360,277){$u_m$ (Hz)}
\put(-353,205){$\theta_m$ ($\pi$)}
\put(-360,134){$u_r$ (Hz)}
\put(-353,62){$\theta_m$ ($\pi$)}
\caption{
The real amplitudes and phases of the microwave and radio-frequency pulses engineered by the GRAPE algorithm to create an entangling gate which is  the robust against the pulse-length errors (PLE). The application time is $6\pi \mu s$ and the maximum real amplitude is 1 MHz.
\label{Grape_Pul}}
\end{figure}

\begin{figure}[ht]
\includegraphics[scale=0.6]{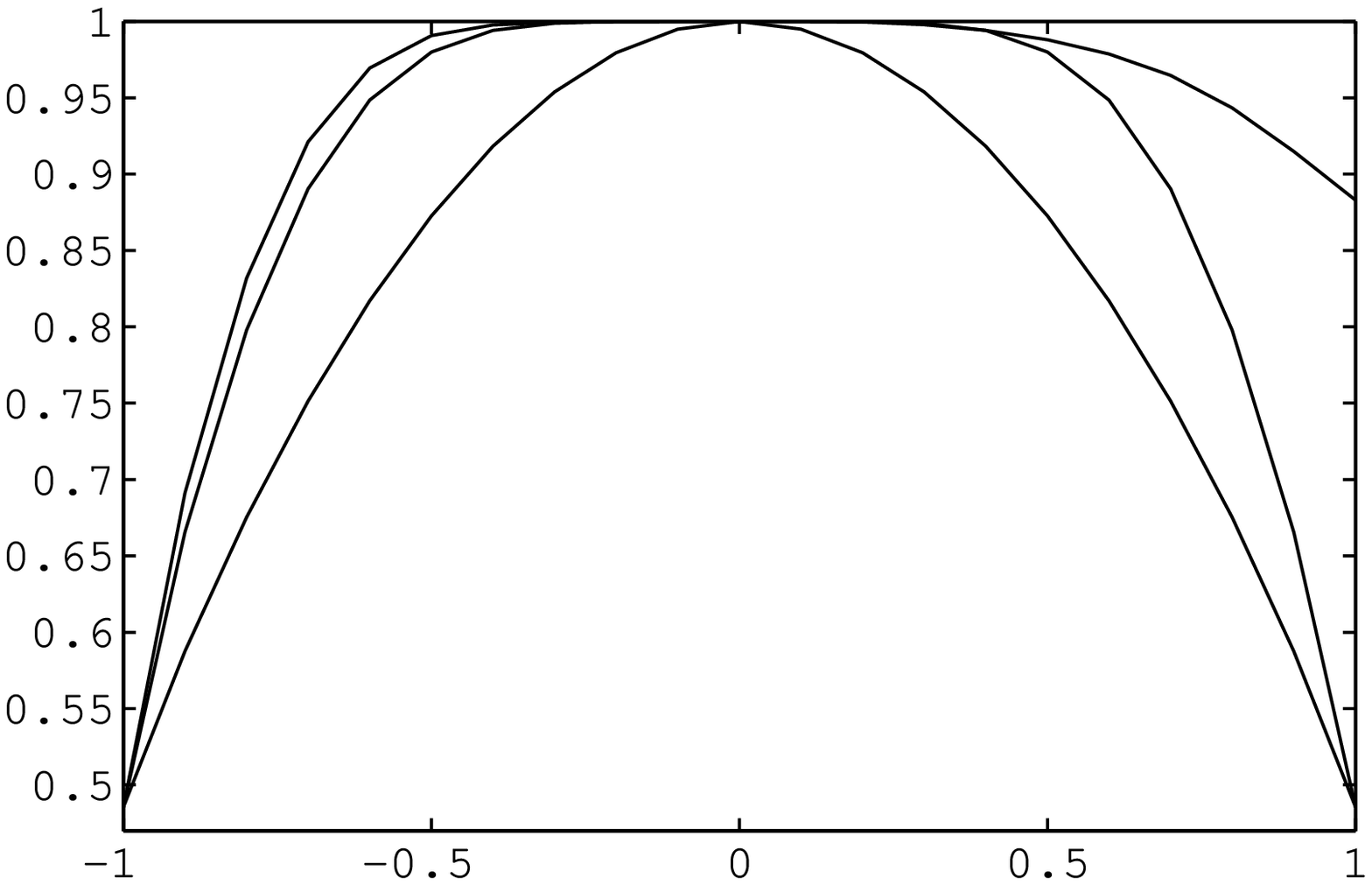}
\put(-230,-5){Pulse-length error fractions $\epsilon_f$}
\put(-320,100){$F$}
\put(-75,100){1}
\put(-70,130){2}
\put(-60,160){3}
\put(-242,100){1}
\put(-250,130){2}
\put(-255,160){3}
\caption{
Fidelity plots for the sequential pulse (Line 1), the BB1 composite pulse (Line 2) and the GRAPE pulse (Line 3) against PLE.
\label{Fidelity_Pul}}
\end{figure}

\begin{figure}[ht]
\includegraphics[scale=0.65]{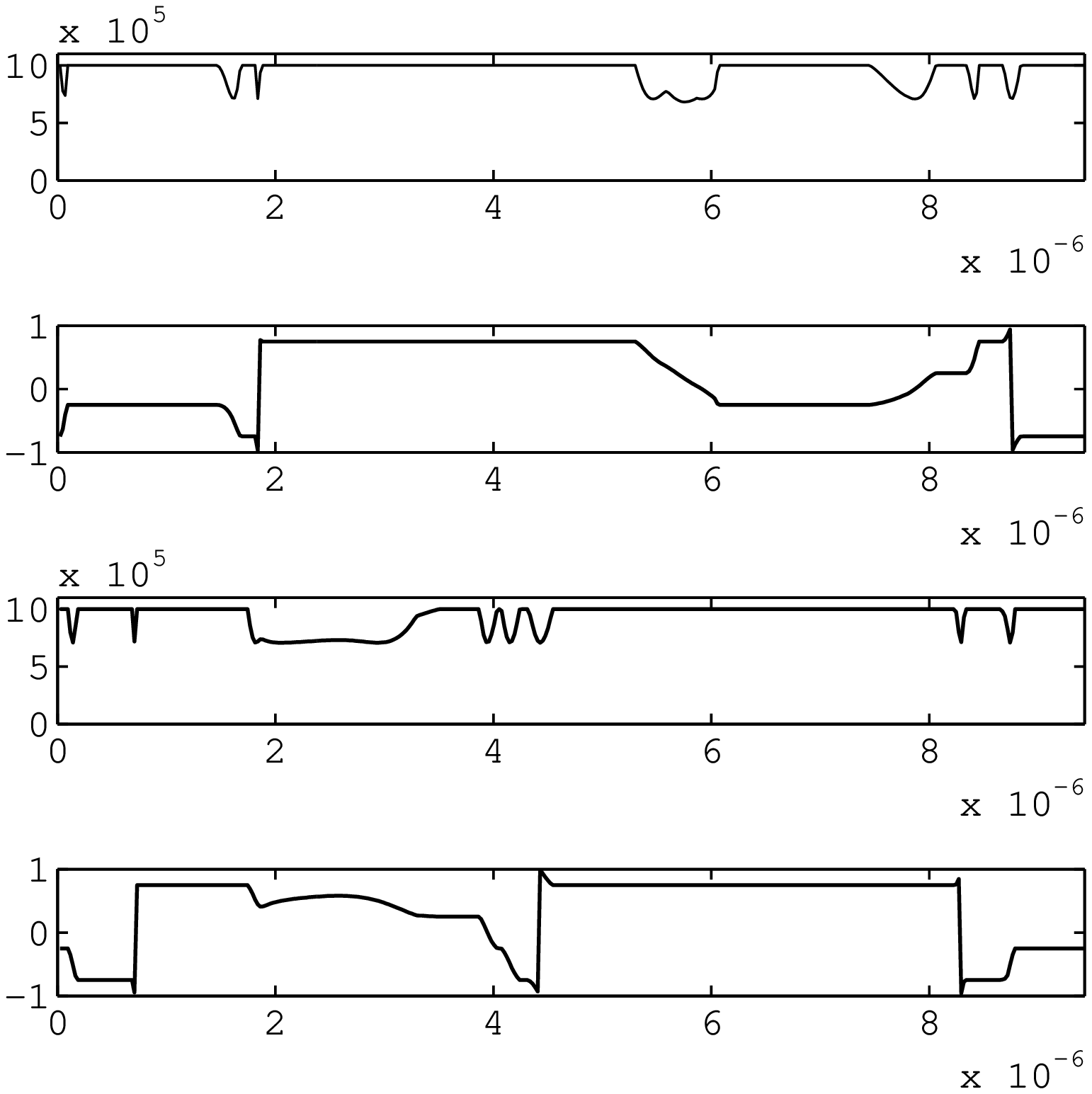}
\put(-180,10){Time (s)}
\put(-360,277){$u_m$ (Hz)}
\put(-353,205){$\theta_m$ ($\pi$)}
\put(-360,134){$u_r$ (Hz)}
\put(-353,62){$\theta_m$ ($\pi$)}
\caption{
The real amplitude and phases of microwave and radio-frequency pulses engineered by the GRAPE algorithm to create an entangling gate which is  the robust against off-resonance errors (ORE). The application time and the maximum real amplitude are the same as those in the case of PLE.
\label{Grape_Off}}
\end{figure}

\begin{figure}[ht]
\includegraphics[scale=0.6]{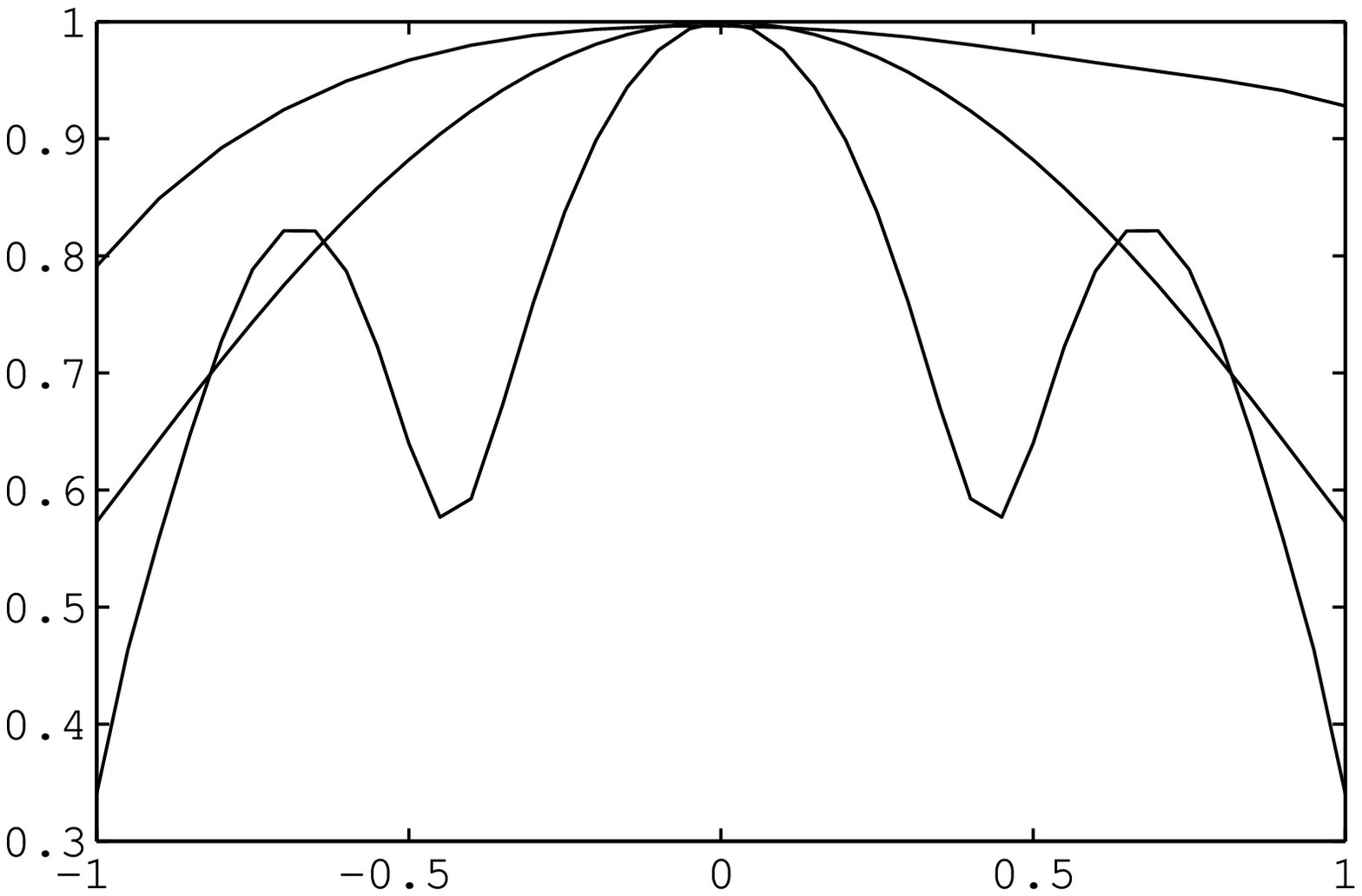}
\put(-230,-5){Off-resonance error fractions $\epsilon_g$}
\put(-320,100){$F$}
\put(-90,100){2}
\put(-90,140){1}
\put(-90,168){3}
\put(-227,100){2}
\put(-227,140){1}
\put(-227,168){3}
\caption{
Fidelity plots for the sequential pulse (Line 1), the CORPSE composite pulse (Line 2) and the GRAPE pulse (Line 3) against ORE.
\label{Fidelity_Off}}
\end{figure}

In our numerical simulations, the modified GRAPE algorithms explained in the previous section are executed using Matlab to produce a set of optimised microwave and radio-frequency pulses presented in Figure 2 and 4. Both sets of pulses have $N=400$ and $\max\{u_m\}=\max\{u_r\} = 1$  MHz and this limit is realistic in current NV ESR/NMR ODMR experiments. We found that by simultaneous irradiations we are able to reduce the time taken to achieve robust operation with respect to both the pulse-length and off-resonance errors. We manually re-optimise the pulses by adjusting the range of the error fractions and the duration several times until the gate reaches a considerably high fidelity.

In both cases of pulse-length and off-resonance errors, the gate fidelity of the optimised pulses via GRAPE outperforms those of the sequential pulses and the composite pulses for every value of the PLE and ORE fractions, as shown in Figure 3 and 5. However, the BB1 composite pulse does better than the sequential pulse in the case of pulse-length error. It is interesting to note that the CORPSE composite pulse, which is robust against ORE in a single two-level system, is no longer robust against the ORE in our essentially three-level system. Hence, a further investigations would be needed in order to develop composite pulses that can tackle the off-resonance error in a three-level system. This is beyond the scope of this paper. 

While the maximum real amplitudes of the optimised pulses used in our analyses are not technologically difficult to implement  \cite{exper1,exper2,experGHZ}, the fast controlling apparatuses required to rapidly change the pulse amplitudes and phases is quite demanding since they should be able to operate on nanosecond timescales.

In conclusions, we have numerically optimised the microwave and radio-frequency pulses required to create the entanglement in the system of single NV centre coupled to the nearest Carbon atom through the modified GRAPE algorithm. We have found that the entangling gate created by such optimised pulses is more robust against systematic errors and has faster implementation time than that required by the corresponding composite pulses. Reasonable extensions of our work would be considering the simultaneous presence of PLE and ORE in the system and taking into account decoherence processes. The latter one certainly needs the system to be modelled by an open quantum system analysis through master equation.  

This work was supported under the European Commission FP6 IST FET QIPC project QAP Contract Number 015848.


\newpage
\begin{thebibliography}{99}
\bibitem{solidQC}
B. E. Kane, Prog. in Phys., {\bf 48} 1023 (2000).
\bibitem{DonorQC}
J. J. L. Morton, et. al., Nature {\bf455} 1085 (2008).
\bibitem{QuantumDotsQC}
M. Kroutvar, et. al., Nature {\bf432} 81 (2004).
\bibitem{Nakamura}
Y. Nakamura, et. al., Nature {\bf398} 786 (1999).
\bibitem{cQED}
R. J. Schoelkopf and S. M. Girvin, Nature {\bf451} 664 (2008).
\bibitem{NVScience}
M.V.G. Dutt, et. al., Science {\bf316} 1312 (2007).
\bibitem{NVQCreview}
F. Jelezko and J. Wrachtrup, J. Phys. Condens. Matter {\bf 16} R1089 (2004). 
\bibitem{exper1}
F. Jelezko, et. al., Phys. Rev. Lett. {\bf 92} 076401 (2004)
\bibitem{exper2}
F. Jelezko, et. al., Phys. Rev. Lett. {\bf 93} 130501 (2004).
\bibitem{AP}
A. P. Nizovtsev, et. al., Optics and Spectrosc. {\bf 99} 248 (2005).
\bibitem{experGHZ}
P. Neumann, et. al., Science {\bf 320} 01326 (2008).
\bibitem{grape}
N. Khaneja, et. al., J. Magn. Reson. {\bf 172} 296 (2005).
\bibitem{composite}
H. K. Cummins, et. al., Phys. Rev. A {\bf 67} 042308 (2003).
\bibitem{grape2}
T. Schulte-Herbr\"{u}ggen et. al., Phys. Rev. A {\bf 72} 042331 (2005).
\bibitem{Levitt}
M. H. Levitt and R. Freeman, J. Magn. Reson. {\bf33} 473 (1979).
\bibitem{compositeNJP}
H. K. Cummins and J. A. Jones, New. J. Phys. {\bf2} 6.1 (2000).
\bibitem{Wei1999}C. Wei and N. B. Manson, Phys. Rev. A {\bf 60}, 2540 (1999).
\bibitem{ion}
N. Timoney, et. al., Phys. Rev. A {\bf 77} 052334 (2008).
\bibitem{Josephson}
A. Sp\"{o}rl, et. al., Phys. Rev. A {\bf 75} 012302 (2007).
\bibitem{Markov}
T. Schulte-Herbr\"{u}ggen et. al., e-print arXiv:quant-ph/0609037 (2006).
\bibitem{Mikko}
M. M\"ott\"onen, et. al., Phys. Rev. A {\bf 73} 022332 (2006).
\end{thebibliography}
\end{document}